\documentclass[pra,twocolumn,showpacs,preprintnumbers,amsmath,amssymb]{revtex4}

\usepackage{graphicx}
\usepackage{dcolumn}
\usepackage{bm}

\begin{document}

\title{
Theory of ``magic'' optical traps for Zeeman-insensitive clock transitions in alkalis
}
\author{Andrei Derevianko}
\email{andrei@unr.edu}
\affiliation{Department of Physics, University of Nevada, Reno NV 89557}

\date{\today}

\begin{abstract}
Precision measurements and quantum information processing with cold atoms may benefit
from trapping atoms with specially engineered, ``magic'' optical fields.   At
the magic trapping conditions, the relevant atomic properties remain immune to
strong perturbations by the trapping fields. Here we develop a theoretical
analysis of magic trapping  for especially valuable Zeeman-insensitive
clock  transitions in alkali-metal atoms. The involved mechanism relies on applying
``magic'' bias B-field along a circularly polarized trapping laser field.
We map out these B-fields as a function of trapping laser wavelength
for all commonly-used alkalis. We also highlight a common error in evaluating
Stark shifts of hyperfine manifolds.
\end{abstract}





\maketitle

A recurring theme in modern precision measurements and quantum information
processing with cold atoms and molecules are the so-called ``magic'' traps~\cite{YeKimKat08}. At
the magic trapping conditions, the relevant atomic properties remain immune to
strong perturbations by optical trapping fields. For example, in optical
lattice clocks, the atoms are held using laser fields operating at magic
wavelengths~\cite{KatTakPal03}. The clock levels are shifted due to the dynamic Stark effect that
depends on the trapping laser wavelength. At the specially-chosen, ``magic'',
wavelength, both clock levels are perturbed identically; therefore the
differential effect of trapping fields simply vanishes for the clock transition. This
turned out to be a powerful idea: lattice clocks based on the alkaline-earth
atom Sr have recently outperformed the primary frequency standards~\cite{LudZelCam08}.

Finding similar magic conditions for ubiquitous alkali-metal atoms employed in
a majority of cold-atom experiments remains an open challenge. Especially
valuable are the microwave transitions in the ground-state hyperfine manifold. Finding magic conditions here, for example, would enable
developing microMagic clocks~\cite{BelDerDzu08Clock}:
microwave clocks with the active clockwork area
of a few micrometers across.
In addition, the hyperfine manifolds are used to
store quantum information in a large fraction of quantum computing
proposals with ultracold alkalis. Here the strong
perturbation due to trapping fields is detrimental. Namely the dynamic
differential Stark shifts is the limiting experimental factor for realizing long-lived quantum
memory~\cite{ZhaDudJen09}. Qualitatively, as an atom moves in the trap, it randomly samples
various intensities of the laser field; this leads to an accumulation of
uncontrolled phase difference between the two qubit states. For very cold samples,
the accumulation of uncontrolled phases may arise because the interrogation by the microwaves is an ensemble average over the spatial distribution of atoms across the trap.
Magic conditions rectify these
problems, as both qubit states see the very same optical potential and do not
accumulate differential phase at all. In other words, we engineer a decoherence-free trap.

Initial steps in identifying magic conditions for hyperfine transitions in
alkali-metal atoms have been made in Refs.~\cite{RosGheDzu09,FlaDzuDer08,ChoCho07}.
The proposals~\cite{FlaDzuDer08,ChoCho07} identified
magic conditions for $M_{F}\neq0$ states. Due to non-vanishing magnetic
moments, these states, however, are sensitive to stray magnetic fields which
would lead to clock inaccuracies and decoherences (except for special cases of relatively large bias fields, see below). Recently, it has been
realized by Lundblad et al.\cite{LunSchPor09} that magic conditions may be attained for the Zeeman-insensitive $M_{F}=0$ states as well.
Here the bias magnetic field is tuned to make the conditions
``magic'' for a given trapping laser wavelength. These authors
experimentally demonstrated these conditions for lattice-confined Rb atoms at a single wavelength.
As demonstrated below, mapping out values of magic bias B-fields for a wide range of
wavelengths requires full-scale structure calculations.
Below I carry out such calculations and point out common pitfalls in
evaluating differential polarizabilities of hyperfine manifolds.

In this work, we are interested in the clock transition  of frequency $\nu_{0}$ between two hyperfine
states $\left\vert F^{\prime}=I+1/2,M_{F}^{\prime}=0\right\rangle $ and
$\left\vert F=I-1/2,M_{F}=0\right\rangle$ attached to the ground electronic $nS_{1/2}$ state of an alkali-metal atom ($I$ is the nuclear spin). Here and below we  denote
the upper clock state as $|F'\rangle$ and the lower state as $|F\rangle$. The magic conditions
are defined as the clock frequency being independent on the perturbing trapping optical field.

We start by reviewing the Zeeman effect for the
clock states. The Zeeman Hamiltonian reads $H^{Z}=-\mu_{z}B$, $\mu$ being the magnetic
moment operator. The permanent magnetic moments of the $M_{F}=0$ states vanish,
so the effect arises in the second order.
We need to diagonalize the following Hamiltonian%
\begin{equation}
H_{\mathrm{eff}}^{Z}=\left(
\begin{tabular}
[c]{ll}%
$h\nu_{0}$ & $H_{F^{\prime}F}^{Z}$\\
$H_{FF^{\prime}}^{Z}$ & $0$%
\end{tabular}
\ \right)  . \label{Eq:HeffZ}%
\end{equation}
The leading effect is due to off-diagonal
coupling $H_{FF^{\prime}}^{Z}=\langle F^{\prime},M_{F}^{\prime}%
=0|H^{Z}|F,M_{F}=0\rangle$.
In case of alkalis, $\left(  \mu_{z}\right)
_{FF^{\prime}}\approx\mu_{B}$, where $\mu_{B}$ is the Bohr magneton. The
resulting Zeeman substates repeal each other and in sufficiently weak
B-fields, $\mu_B B \ll h\nu_{0}$,  the shift of the transition frequency is quadratic in magnetic field,
\begin{equation}
\frac{\delta\nu_{Z}\left(  B\right)  }{\nu_{0}}\approx2\left(  \frac{\mu_{B}%
}{h\nu_{0}}B\right)  ^{2}.
\end{equation}


Since  atoms are trapped by a laser field, the atomic levels are shifted
due to the dynamic Stark effect (see, e.g., a review~\cite{ManOvsRap86}).
The relevant energy-shift operator reads
\[
\hat{U}\left(  \omega_{L}\right)  =-\hat{\alpha}\left(  \omega_{L}\right)
~\left(  \frac{E_{L}}{2}\right)  ^{2},
\]
where $E_L$ is the amplitude of the laser field and $\hat{\alpha}\left(  \omega_{L}\right)$
is the operator of dynamic atomic polarizability; it depends on the laser frequency.
Notice
that $\hat{U}$ may have both diagonal and off-diagonal matrix elements between
atomic states of the same parity.

Now we add the Stark shift couplings to the Hamiltonian (\ref{Eq:HeffZ}). The
Stark shift operator has both the diagonal and off-diagonal matrix elements in the
clock basis. To find the perturbed energy levels, we diagonalize the effective
Hamiltonian%
\begin{equation}
H_{\mathrm{eff}}=\left(
\begin{tabular}
[c]{ll}%
$h\nu_{0}+U_{F^{\prime}F^{\prime}}$ & $U_{F^{\prime}F}+H_{F^{\prime}F}^{Z}$\\
$U_{FF^{\prime}}+H_{FF^{\prime}}^{Z}$ & $U_{FF}$%
\end{tabular}
\right)  .
\end{equation}
For sufficiently weak fields, the resulting shift of the clock frequency
reads
\begin{equation}
\delta\nu_{\mathrm{clock}}\left(  \omega_{L},B,E_{L}\right)  =\delta\nu
_{Z}\left(  B\right)  +\delta\nu_{S}\left(  \omega_{L},B,E_{L}\right)
\end{equation}
with the Stark shift
\begin{align}
&\delta\nu_{S}\left(  \omega_{L},B,E_{L}\right) =  \frac{1}{h}\left\{
\alpha_{F^{\prime}F^{\prime}}\left(  \omega_{L}\right)  -\alpha_{FF}\left(
\omega_{L}\right)   \right. \nonumber \\
&  \left.  -\left(  \frac{4\mu_{FF^{\prime}}B}{h\nu_{0}}\right)
\alpha_{F^{\prime}F}\left(  \omega_{L}\right)  \right\}  \left(  \frac{E_{L}%
}{2}\right) ^{2} . \label{Eq:StarkShift}
\end{align}
The ``magic'' conditions are attained when $\delta\nu_{S}\left(  \omega
_{L},B,E_{L}\right) =0$ for any value of the laser amplitude, i.e., simply
when the combination inside the curly brackets vanishes.

At this point one may evaluate the dynamic polarizabilities and deduce the magic B-field.
Before proceeding with the analysis, I would like to address common pitfalls in evaluating
polarizabilities of hyperfine-manifold states, so the reader appreciates
the necessity of full-scale calculations. A generic expression for the polarizability of $|nFM_{F}\rangle$ state reads
\begin{equation}
\alpha_{FF}^{\left( 0\right) }\left( \omega\right) =\sum_{i=|n_{i}F_{i}M_{i}\rangle }\frac{\langle nFM_{F}|~D_{z}|i\rangle \langle i~|D_{z}|nFM_{F}\rangle }{E_{nFM_{F}}-E_{i}+\omega} +...
\end{equation}
where the omitted term differs by $\omega \rightarrow -\omega$, and $D$ is the dipole operator. All the involved states are the hyperfine
states. While this requires that  the energies include hyperfine splittings, it also  means
that the wave-functions incorporate hyperfine interaction (HFI) to all-orders of perturbation theory.
Including the experimentally-known hyperfine splittings in the
summations is straightforward and unsophisticated practitioners stop at that,
completely neglecting the HFI corrections to the wave-functions. This is hardly justified as
both contributions are of the same order.

\begin{figure}[h]
\begin{center}
\includegraphics*[scale=1.0]{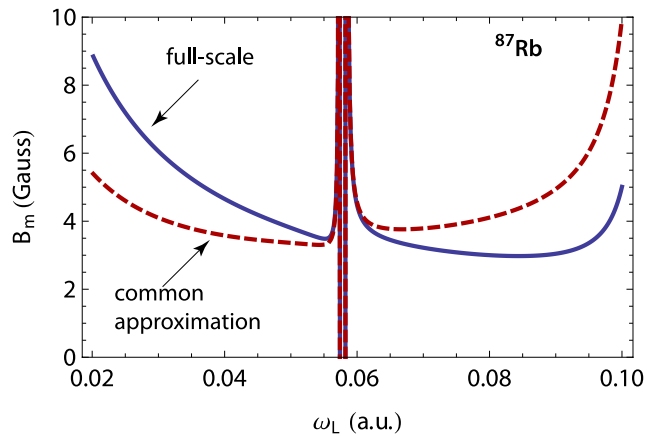}
\end{center}
\caption{(Color online) Importance of full-scale calculations.
Dependence of magic B-field (in Gauss) on
laser frequency (in atomic units) for
$^{87}$Rb.  Full-scale calculations (solid blue line) are compared with
approximate ``experimentalist'' computations which neglect the
HFI contribution to atomic wave-functions (dashed red line).
 \label{Fig:RbBmagicDumbedDownPlots}}
\end{figure}

I would like to remind the reader of a recent controversy: neglecting the HFI correction to wave-functions  has already lead to a (even qualitatively) wrong identification of magic conditions. The authors of Ref.\cite{ZhoCheChe05} employed the simplified approach and (for $B=0$) found a multitude of magic wavelengths for clock transitions in Cs. The prediction was in a contradiction with a subsequent fountain clock measurement; the full-scale calculations have found that in fact there are no magic wavelengths at $B=0$, Ref.~\cite{RosGheDzu09}. To reinforce this point in the context of this paper,
in Fig.~\ref{Fig:RbBmagicDumbedDownPlots},
I compare  results of two calculations of magic B-fields for $^{87}$Rb as a function of laser frequency. In the first calculation, I  neglected the HFI correction to the wave-functions (while including the hyperfine corrections to the energies), and the second result comes the full-scale calculation described below. We clearly see that the simplified approach is off by as much as a factor of two. Only near the resonance the two approaches produce similar results.

These two examples should convince the reader that the full-scale calculations are indeed required for reliably predicting magic fields.
A consistent approach to evaluating
dynamic polarizabilities of hyperfine states was developed in Ref.\cite{RosGheDzu09}.
The HFI correction to wave-functions and energies was included to the leading order;
this leads to a third-order analysis
quadratic in dipole couplings and linear in the HFI.  Below I simplify the magic field conditions
using the formalism of Ref.\cite{RosGheDzu09}.

We may decompose the  polarizability into a
sum over 0-, 1-, and 2-rank tensors
\begin{equation}
\hat{\alpha}\left(  \omega_{L}\right)=
 \hat{\alpha}^{\left(  0\right)  }\left(  \omega_{L}\right)
+A \, \hat{\alpha}^{\left(  1\right)  }\left(  \omega_{L}\right)
+\hat{\alpha}^{\left(  2\right)  }\left(  \omega_{L}\right)  \, .
\end{equation}
These terms are conventionally referred to as the scalar, vector (axial), and
tensor contributions.
 We also
explicitly factored out the degree of circular polarization $A$ of the wave
($A=\pm1$ for pure $\sigma_{\pm}$ light). The direction of the bias B-field defines the
quantization axis. We also fixed the direction of the wave propagation
$\mathbf{\hat{k}}$ to be parallel to the B-field. Notice that the circular
polarization of the optical field is defined with respect to the quantization
axis (not $\mathbf{\hat{k}}$).

Below we show that the ``magic'' value of the magnetic field may be represented as
\begin{equation}
B_{m}\left(  \omega_{L}\right)  \approx-\frac{1}{\mu_{B}}\frac{2I+1}{2I}%
\frac{\alpha_{FF}^{\left(  0\right)  ,\mathrm{HFI}}\left(  \omega_{L}\right)
}{A\alpha_{nS_{1/2}}^{a}\left(  \omega_{L}\right)  }~h\nu_{0}
.\label{Eg:MagicB}
\end{equation}
It depends on the laser frequency and the degree of circular polarization $A$, $|A| \le 1$.
$\alpha_{FF}^{\left(  0\right)  ,\mathrm{HFI}}\left(  \omega_{L}\right)$ is the scalar
HFI-mediated third-order polarizability of the lower clock state, $F=I-1/2$.


Indeed, the non-vanishing contribution to the differential polarizability
$\Delta\alpha\left(  \omega_{L}\right)  =\alpha_{F^{\prime}F^{\prime}}\left(
\omega_{L}\right)  -\alpha_{FF}\left(  \omega_{L}\right)$, entering Eq.~(\ref{Eq:StarkShift}), comes only through
the hyperfine-mediated interactions:
$\Delta\alpha (  \omega_{L} )  =
\alpha_{F^{\prime}F^{\prime}}^{\mathrm{HFI}}\left(  \omega_{L}\right)  -\alpha_{FF}^{\mathrm{HFI}}\left(
\omega_{L}\right)$. This reflects the fact that both hyperfine levels belong
to the same electronic configuration - the symmetry in responding to fields is
only broken when the HFI is included. Moreover, for alkalis $\alpha_{FF}$ and
$\alpha_{F^{\prime}F^{\prime}}$ are dominated by the \emph{scalar} part of
polarizability: $\Delta\alpha\left(  \omega_{L}\right)  \approx \alpha_{F^{\prime
}F^{\prime}}^{\left(  0\right)  ,\mathrm{HFI}}\left(  \omega_{L}\right)
-\alpha_{FF}^{\left(  0\right)  ,\mathrm{HFI}}\left(  \omega_{L}\right)  $.
These two polarizabilities never intersect -- they are strictly proportional
to each other:
$\alpha_{F^{\prime}F^{\prime}}^{\left(  0\right),\mathrm{HFI}}\left(  \omega_{L}\right)
=-(I+1)/I~\alpha_{FF}^{\left(0\right)  ,\mathrm{HFI}}\left(  \omega_{L}\right)$.

Now we turn to simplifying the
off-diagonal matrix element $\alpha_{F^{\prime}F}\left(  \omega_{L}\right)$ entering 
Eq.~(\ref{Eq:StarkShift}). It
is dominated by the vector part of polarizability. Indeed, $\langle
F^{\prime},M_{F}^{\prime}|\hat{\alpha}^{\left(  0\right)  }|F,M_{F}\rangle=0$
due to the angular selection rules ($F^{\prime}\neq F$). While the tensor
contribution $\langle F^{\prime},M_{F}^{\prime}|\hat{\alpha}^{\left(
2\right)  }|F,M_{F}\rangle$ does not vanish, the electronic momentum of the
ground state $nS_{1/2}$ is $J=1/2$; therefore (since $\langle J=1/2|\hat{\alpha}^{\left(
2\right)  }|J=1/2\rangle \equiv 0$)
  this matrix element requires the HFI
admixture and becomes strongly suppressed. By contrast, the vector
contribution $\langle F^{\prime},M_{F}^{\prime}|\hat{\alpha}^{\left(
1\right)  }|F,M_{F}\rangle~$\ does not vanish even if the hyperfine couplings
are neglected. It is worth mentioning that it arises only due to relativistic effects, since the orbital
angular momentum $L=0$ for the ground state; e.g., vector polarizability is much smaller in Li than in Cs.
The off-diagonal matrix element
of the rank-1 polarizability may be expressed as $\alpha_{F^{\prime}%
F}^{\left(  1\right)  }\left(  \omega_{L}\right)  =\frac{1}{2}\alpha
_{nS_{1/2}}^{a}\left(  \omega_{L}\right)  $, where $\alpha_{J}^{a}\left(
\omega_{L}\right) $ is the conventionally-defined second-order vector
polarizability of the ground $nS_{1/2}$ state.

To evaluate the polarizabilities, we used a blend of relativistic many-body
techniques of atomic structure, as described in~\cite{BelSafDer06}.
To improve upon the accuracy, high-precision experimental data
were used where available. To ensure the quality of the calculations, a
comparison with the experimental literature data on static Stark shifts of the
clock transitions was made. Overall, we expect the theoretical errors not to exceed
1\% for Cs and to be at the level of a few 0.1\% for lighter alkalis.
If required, better accuracies may be reached with many-body methods developed for
analyzing atomic parity violation~\cite{PorBelDer09}.

\begin{figure}[h]
\begin{center}
\includegraphics*[scale=0.6]{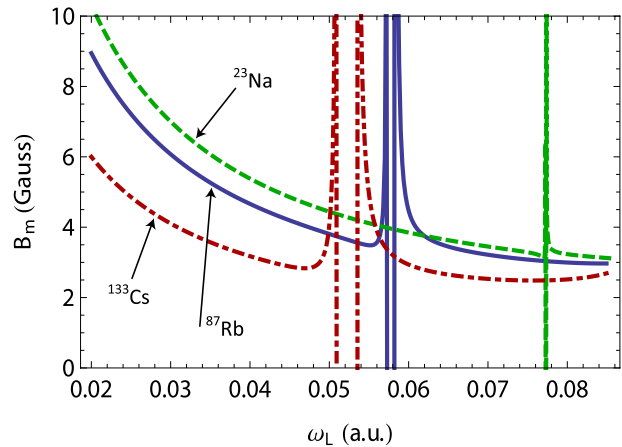}
\end{center}
\caption{(Color online) Dependence of magic B-field (in Gauss) on
laser frequency (in atomic units) for $^{23}$Na (dashed green  line),
$^{87}$Rb (solid blue line), and $^{133}$Cs (dot-dashed red  line).
 Magic B-fields for other isotopes of the same element may be obtained  using the scaling law, Eq.~(\ref{Eq:IsoScaling}).
 \label{Fig:BmagicPlots}}
\end{figure}

Our computed dependence of magic B-field on laser frequency for
representative alkalis ($^{23}$Na,$^{87}$Rb, and $^{133}$Cs)   is shown in Fig.~\ref{Fig:BmagicPlots}. We also carried out similar calculations for
$^{39}$K and $^7$Li. Results for several laser wavelengths are presented in Table~\ref{Tab:atoms}.


\begin{table}[h]
\caption{ Values of magic B-fields for representative laser wavelengths. The optical
field is assumed to be purely circularly polarized. Values of the clock transition frequencies $\nu_0$ and the second-order Zeeman frequency shift coefficients $\delta\nu_Z/B^2$ are listed in the
second and the third columns, respectively. Magic B-fields for other isotopes of the same element may be obtained with the scaling law, Eq.~(\ref{Eq:IsoScaling}).
 \label{Tab:atoms}}
\begin{ruledtabular}
\begin{tabular}{clcccc}
  &  $\nu_{0}$ &  $\delta\nu_Z/B^2$ & \multicolumn{3}{c}{``magic'' B (Gauss)} \\
  &   (GHz)    &   (kHz/G$^2$)       &
          10.6 $\mu$m & 1.065 $\mu$m &  811.5 nm     \\
\hline
$^{7}$Li   &  0.80  & 4.9  &   -  &  144  & 64.9   \\
$^{23}$Na  &  1.77  & 2.2  & 47.4 &  5.07 & 4.05   \\
$^{39}$K   &  0.46  & 8.5  & 0.782& 0.0848& 0.0672 \\
$^{87}$Rb  &  6.83  & 0.57 & 41.0 &  4.39 & 3.62       \\
$^{133}$Cs &  9.19  & 0.43 & 27.3 &  3.00 & 3.81        \\
\end{tabular}
\end{ruledtabular}
\end{table}

From Fig.~\ref{Fig:BmagicPlots} we observe that below the resonances, magic B-fields grow smaller with increasing
laser frequency. This is a reflection of the fact that at small $\omega_L$, the HFI-mediated polarizability
approaches a constant value, while the vector polarizability $\propto \omega_L$. Thus, $B_m \propto 1/\omega_L$
in accord with Fig.~\ref{Fig:BmagicPlots}. As the frequency is increased, the $B_m(\omega_L)$ increases near the atomic resonance (fine-structure doublet). This leads to a prominent elbow-like minimum in the $B_m(\omega_L)$ curves.

Magic B-field has been recently measured for optical lattice-confined $^{87}$Rb at 811.5 nm, Ref.~\cite{LunSchPor09}. At this wavelength and degree of circular polarization $A=0.77(4)$, the measured $B_m = 4.314(3)\, \mathrm{Gauss}$. For a purely circularly-polarized light, this translates into $B_m=3.32(17)\, \mathrm{Gauss}$.
The computed magic B-field, $3.62 \, \mathrm{Gauss}$, is $1.8 \sigma$ larger than the measured value.

A quick glance through the Table~\ref{Tab:atoms} reveals that the required B-fields for $^{39}$K are
much weaker than for other alkalis; this is related to the fact that the nuclear moment of this isotope is
almost an order of magnitude smaller than that of other species. An additional suppression is due to the magic B-fields being {\em quadratic} in hyperfine splitting (clock frequency).

Notice that if the B-field and the direction of laser propagation are set at an angle $\theta$,
then $A \rightarrow A \cos \theta$ in Eq.(\ref{Eg:MagicB}) (see Ref.~\cite{FlaDzuDer08}). This angle provides an additional
experimental handle on reaching the magic conditions. Increasing the angle and reducing
the degree of circular polarization raise values of the magic B-field.

Generically, the ratio $\alpha_{FF}^{\left(  0\right)  ,\mathrm{HFI}}\left(
\omega_{L}\right)  /\alpha_{nS_{1/2}}^{a}\left(  \omega_{L}\right)  $ is in
the order of a ratio of the hyperfine splitting to the fine-structure
splitting in the nearest $P$-state manifold, i.e., it is much smaller than
unity. This reinforces the validity of the weak-field approximation used to
derive Eqs.(\ref{Eq:StarkShift},\ref{Eg:MagicB}). Notice, however, that
$\lim_{\omega_{L}\rightarrow0}\alpha_{nS_{1/2}}^{a}\left(  \omega_{L}\right)
\rightarrow 0$; this may lead to unreasonably large magic
B-fields for very low-frequency fields. Such a breakdown occurs for $^7$Li
at 10.6 $\mu$m in Table~\ref{Tab:atoms}.

It is worth pointing out that the results of Fig.~\ref{Fig:BmagicPlots} and Table~\ref{Tab:atoms}
may be extended to other, e.g., unstable, isotopes. An analysis of the
third-order expressions for the HFI-mediated polarizabilities shows that the
magic B-fields scales with the nuclear spin and g-factor as
\begin{equation}
B_{m}\propto g_{I}^{2}\frac{2I~\left( 2I+1\right) ^{2}}{\left( 2I+2\right) ^{3/2}}
\label{Eq:IsoScaling} \, .
\end{equation}

Finally, I would like to comment on the magic conditions for the $M_F\neq 0$ states discussed in our earlier work~\cite{FlaDzuDer08}. The idea there was to rotate the bias B-field with respect to the laser propagation. At a certain laser-frequency-dependent magic ``angle'', $\theta\approx 90^\circ$, contributions of the HFI-meditated scalar polarizability and the rotationally-suppressed vector polarizability were compensating
each other. Notice that we may attain the Zeeman-insensitivity even in this case. Indeed, in magnetic field, two hyperfine levels $|F=I+1/2,M_F\rangle$ and $|F=I-1/2,M_F\rangle$ repel each other through off-diagonal Zeeman coupling. In addition, the g-factors of the two levels have opposite signs. This leads to a minimum in the clock-frequency dependence on B-fields. These minima occur at relatively large magnetic fields, e.g., about 2 kGauss for $^{87}$Rb. This $d \nu(B)/dB=0$ condition fixes ``magic'' B-field values for the proposal~\cite{FlaDzuDer08}.

It is anticipated that a variety of applications could take advantage of the magic conditions computed in this paper. For example, the dynamic Stark shift is the primary factor limiting lifetime of quantum memory~\cite{ZhaDudJen09}; here an advance may be made by switching to the magic B-fields.
It remains to be seen if the microMagic lattice clock can be developed; here one needs to investigate the feasibility of stabilizing bias magnetic fields at the magic values. In this regard, notice that we still have a choice of fixing
laser wavelength/polarization/rotation angle to optimize clock accuracy with respect to drifts in the B-field.

\emph{Acknowledgements ---}
 I would like to thank Trey Porto, Nathan Lundblad, and Alex Kuzmich for discussions.  This work was supported in part by the US NSF and by the US NASA under Grant/Cooperative Agreement No. NNX07AT65A issued by the Nevada NASA EPSCoR program.


\begin{thebibliography}{13}
\expandafter\ifx\csname natexlab\endcsname\relax\def\natexlab#1{#1}\fi
\expandafter\ifx\csname bibnamefont\endcsname\relax
  \def\bibnamefont#1{#1}\fi
\expandafter\ifx\csname bibfnamefont\endcsname\relax
  \def\bibfnamefont#1{#1}\fi
\expandafter\ifx\csname citenamefont\endcsname\relax
  \def\citenamefont#1{#1}\fi
\expandafter\ifx\csname url\endcsname\relax
  \def\url#1{\texttt{#1}}\fi
\expandafter\ifx\csname urlprefix\endcsname\relax\def\urlprefix{URL }\fi
\providecommand{\bibinfo}[2]{#2}
\providecommand{\eprint}[2][]{\url{#2}}

\bibitem[{\citenamefont{Ye et~al.}(2008)\citenamefont{Ye, Kimble, and
  Katori}}]{YeKimKat08}
\bibinfo{author}{\bibfnamefont{J.}~\bibnamefont{Ye}},
  \bibinfo{author}{\bibfnamefont{H.~J.} \bibnamefont{Kimble}},
  \bibnamefont{and} \bibinfo{author}{\bibfnamefont{H.}~\bibnamefont{Katori}},
  \bibinfo{journal}{Science} \textbf{\bibinfo{volume}{320}},
  \bibinfo{pages}{1734} (\bibinfo{year}{2008}).

\bibitem[{\citenamefont{Katori et~al.}(2003)\citenamefont{Katori, Takamoto,
  Pal'chikov, and Ovsiannikov}}]{KatTakPal03}
\bibinfo{author}{\bibfnamefont{H.}~\bibnamefont{Katori}},
  \bibinfo{author}{\bibfnamefont{M.}~\bibnamefont{Takamoto}},
  \bibinfo{author}{\bibfnamefont{V.~G.} \bibnamefont{Pal'chikov}},
  \bibnamefont{and} \bibinfo{author}{\bibfnamefont{V.~D.}
  \bibnamefont{Ovsiannikov}}, \bibinfo{journal}{Phys.\ Rev.\ Lett.}
  \textbf{\bibinfo{volume}{91}}, \bibinfo{pages}{173005}
  (\bibinfo{year}{2003}).

\bibitem[{\citenamefont{Ludlow et~al.}(2008)\citenamefont{Ludlow, Zelevinsky,
  Campbell, Blatt, Boyd, de~Miranda, Martin, Thomsen, Foreman, Ye
  et~al.}}]{LudZelCam08}
\bibinfo{author}{\bibfnamefont{A.~D.} \bibnamefont{Ludlow}},
  \bibinfo{author}{\bibfnamefont{T.}~\bibnamefont{Zelevinsky}},
  \bibinfo{author}{\bibfnamefont{G.~K.} \bibnamefont{Campbell}},
  \bibinfo{author}{\bibfnamefont{S.}~\bibnamefont{Blatt}},
  \bibinfo{author}{\bibfnamefont{M.~M.} \bibnamefont{Boyd}},
  \bibinfo{author}{\bibfnamefont{M.~H.~G.} \bibnamefont{de~Miranda}},
  \bibinfo{author}{\bibfnamefont{M.~J.} \bibnamefont{Martin}},
  \bibinfo{author}{\bibfnamefont{J.~W.} \bibnamefont{Thomsen}},
  \bibinfo{author}{\bibfnamefont{S.~M.} \bibnamefont{Foreman}},
  \bibinfo{author}{\bibfnamefont{J.}~\bibnamefont{Ye}}, \bibnamefont{et~al.},
  \bibinfo{journal}{Science} \textbf{\bibinfo{volume}{319}},
  \bibinfo{pages}{1805} (\bibinfo{year}{2008}).

\bibitem[{\citenamefont{Beloy et~al.}(2009)\citenamefont{Beloy, Derevianko,
  Dzuba, and Flambaum}}]{BelDerDzu08Clock}
\bibinfo{author}{\bibfnamefont{K.}~\bibnamefont{Beloy}},
  \bibinfo{author}{\bibfnamefont{A.}~\bibnamefont{Derevianko}},
  \bibinfo{author}{\bibfnamefont{V.~A.} \bibnamefont{Dzuba}}, \bibnamefont{and}
  \bibinfo{author}{\bibfnamefont{V.~V.} \bibnamefont{Flambaum}},
  \bibinfo{journal}{Phys. Rev. Lett.} \textbf{\bibinfo{volume}{102}},
  \bibinfo{eid}{120801} (pages~\bibinfo{numpages}{4}) (\bibinfo{year}{2009}),
  \urlprefix\url{http://link.aps.org/abstract/PRL/v102/e120801}.

\bibitem[{\citenamefont{Zhao et~al.}(2009)\citenamefont{Zhao, Dudin, Jenkins,
  Campbell, Matsukevich, Kennedy, and Kuzmich}}]{ZhaDudJen09}
\bibinfo{author}{\bibfnamefont{R.}~\bibnamefont{Zhao}},
  \bibinfo{author}{\bibfnamefont{Y.~O.} \bibnamefont{Dudin}},
  \bibinfo{author}{\bibfnamefont{S.~D.} \bibnamefont{Jenkins}},
  \bibinfo{author}{\bibfnamefont{C.~J.} \bibnamefont{Campbell}},
  \bibinfo{author}{\bibfnamefont{D.~N.} \bibnamefont{Matsukevich}},
  \bibinfo{author}{\bibfnamefont{T.~A.~B.} \bibnamefont{Kennedy}},
  \bibnamefont{and} \bibinfo{author}{\bibfnamefont{A.}~\bibnamefont{Kuzmich}},
  \bibinfo{journal}{Nat. Phys} \textbf{\bibinfo{volume}{5}},
  \bibinfo{pages}{100} (\bibinfo{year}{2009}), ISSN \bibinfo{issn}{1745-2473},
  \urlprefix\url{http://dx.doi.org/10.1038/nphys1152}.

\bibitem[{\citenamefont{Rosenbusch et~al.}(2009)\citenamefont{Rosenbusch,
  Ghezali, Dzuba, Flambaum, Beloy, and Derevianko}}]{RosGheDzu09}
\bibinfo{author}{\bibfnamefont{P.}~\bibnamefont{Rosenbusch}},
  \bibinfo{author}{\bibfnamefont{S.}~\bibnamefont{Ghezali}},
  \bibinfo{author}{\bibfnamefont{V.~A.} \bibnamefont{Dzuba}},
  \bibinfo{author}{\bibfnamefont{V.~V.} \bibnamefont{Flambaum}},
  \bibinfo{author}{\bibfnamefont{K.}~\bibnamefont{Beloy}}, \bibnamefont{and}
  \bibinfo{author}{\bibfnamefont{A.}~\bibnamefont{Derevianko}},
  \bibinfo{journal}{Phys. Rev. A} \textbf{\bibinfo{volume}{79}},
  \bibinfo{eid}{013404} (pages~\bibinfo{numpages}{8}) (\bibinfo{year}{2009}),
  \urlprefix\url{http://link.aps.org/abstract/PRA/v79/e013404}.

\bibitem[{\citenamefont{Flambaum et~al.}(2008)\citenamefont{Flambaum, Dzuba,
  and Derevianko}}]{FlaDzuDer08}
\bibinfo{author}{\bibfnamefont{V.~V.} \bibnamefont{Flambaum}},
  \bibinfo{author}{\bibfnamefont{V.~A.} \bibnamefont{Dzuba}}, \bibnamefont{and}
  \bibinfo{author}{\bibfnamefont{A.}~\bibnamefont{Derevianko}},
  \bibinfo{journal}{Phys. Rev. Lett.} \textbf{\bibinfo{volume}{101}},
  \bibinfo{eid}{220801} (\bibinfo{year}{2008}).

\bibitem[{\citenamefont{Choi and Cho}(2007)}]{ChoCho07}
\bibinfo{author}{\bibfnamefont{J.~M.} \bibnamefont{Choi}} \bibnamefont{and}
  \bibinfo{author}{\bibfnamefont{D.}~\bibnamefont{Cho}},
  \bibinfo{journal}{Journal of Physics: Conference Series}
  \textbf{\bibinfo{volume}{80}}, \bibinfo{pages}{012037 (6pp)}
  (\bibinfo{year}{2007}),
  \urlprefix\url{http://stacks.iop.org/1742-6596/80/012037}.

\bibitem[{\citenamefont{Lundblad et~al.}(2009)\citenamefont{Lundblad,
  Schlosser, and Porto}}]{LunSchPor09}
\bibinfo{author}{\bibfnamefont{N.}~\bibnamefont{Lundblad}},
  \bibinfo{author}{\bibfnamefont{M.}~\bibnamefont{Schlosser}},
  \bibnamefont{and} \bibinfo{author}{\bibfnamefont{J.~V.} \bibnamefont{Porto}},
  \emph{\bibinfo{title}{Experimental observation of magic-wavelength behavior
  in optical lattice-trapped $^{87}${R}b}} (\bibinfo{year}{2009}),
  \bibinfo{note}{arXiv.org:0912.1528},
  \urlprefix\url{http://www.citebase.org/abstract?id=oai:arXiv.org:0912.1528}.

\bibitem[{\citenamefont{Manakov et~al.}(1986)\citenamefont{Manakov,
  Ovsiannikov, and Rapoport}}]{ManOvsRap86}
\bibinfo{author}{\bibfnamefont{N.~L.} \bibnamefont{Manakov}},
  \bibinfo{author}{\bibfnamefont{V.~D.} \bibnamefont{Ovsiannikov}},
  \bibnamefont{and} \bibinfo{author}{\bibfnamefont{L.~P.}
  \bibnamefont{Rapoport}}, \bibinfo{journal}{Phys. Rep.}
  \textbf{\bibinfo{volume}{141}}, \bibinfo{pages}{319} (\bibinfo{year}{1986}).

\bibitem[{\citenamefont{Zhou et~al.}(2005)\citenamefont{Zhou, Chen, and
  Chen}}]{ZhoCheChe05}
\bibinfo{author}{\bibfnamefont{X.}~\bibnamefont{Zhou}},
  \bibinfo{author}{\bibfnamefont{X.}~\bibnamefont{Chen}}, \bibnamefont{and}
  \bibinfo{author}{\bibfnamefont{J.}~\bibnamefont{Chen}}
  (\bibinfo{year}{2005}), \eprint{arXiv:0512244}.

\bibitem[{\citenamefont{Beloy et~al.}(2006)\citenamefont{Beloy, Safronova, and
  Derevianko}}]{BelSafDer06}
\bibinfo{author}{\bibfnamefont{K.}~\bibnamefont{Beloy}},
  \bibinfo{author}{\bibfnamefont{U.~I.} \bibnamefont{Safronova}},
  \bibnamefont{and}
  \bibinfo{author}{\bibfnamefont{A.}~\bibnamefont{Derevianko}},
  \bibinfo{journal}{Phys. Rev. Lett.} \textbf{\bibinfo{volume}{97}},
  \bibinfo{pages}{040801} (\bibinfo{year}{2006}).

\bibitem[{\citenamefont{Porsev et~al.}(2009)\citenamefont{Porsev, Beloy, and
  Derevianko}}]{PorBelDer09}
\bibinfo{author}{\bibfnamefont{S.~G.} \bibnamefont{Porsev}},
  \bibinfo{author}{\bibfnamefont{K.}~\bibnamefont{Beloy}}, \bibnamefont{and}
  \bibinfo{author}{\bibfnamefont{A.}~\bibnamefont{Derevianko}},
  \bibinfo{journal}{Phys. Rev. Lett.} \textbf{\bibinfo{volume}{102}},
  \bibinfo{eid}{181601} (pages~\bibinfo{numpages}{4}) (\bibinfo{year}{2009}),
  \urlprefix\url{http://link.aps.org/abstract/PRL/v102/e181601}.

\end{thebibliography}

\end{document}